\documentstyle[twoside,fleqn,espcrc2]{article}

\newcommand{\AmS}{{\protect\the\textfont2
  A\kern-.1667em\lower.5ex\hbox{M}\kern-.125emS}}

\title{The Taming of QCD by Fortran 90
       \thanks{This research is supported in part under DOE grant
               DE-FG02-91ER40676.
               We are grateful to the Center of Computational Science 
               and the Office of Information Technology for support 
               and access to the Boston University supercomputer 
               facility. }}

\author{I.~Dasgupta, A.R.~Levi, V.~Lubicz and C.~Rebbi \\
        Department of Physics, Boston University,
        590 Commonwealth Avenue, Boston, MA 02215, USA}
       
\begin{document}

\begin{abstract}
We implement lattice QCD using the Fortran 90 language. 
We have designed machine independent modules
that define fields (gauge, fermions, scalars, etc...)
and have defined overloaded operators for all possible operations 
between fields, matrices and numbers. 
With these modules it is very simple to write QCD programs. 
We have also created a useful compression standard for 
storing the lattice configurations, 
a parallel implementation of the random generators,
an assignment that does not require temporaries, 
and a machine independent precision definition. 
We have tested our program on parallel and single
processor supercomputers obtaining excellent performances.
\end{abstract}

\maketitle

\section{Introduction}
\label{intro}

With the twofold goal of facilitating the development of algorithms
and applications for lattice QCD,
and of maintaining good code performance, we have taken advantage of
the possibilities offered by Fortran 90 to write a set of modules for
a high-level, yet efficient implementation of QCD simulations.
In particular Fortran 90 offers the possibility to
define both types and overloaded operators.
These two key features make Fortran 90 particularly suitable
for QCD simulations.

Our end product is a package, ``QCDF90'',
which is fully described in a long documentation \cite{QCDF90},
where we provide all the information needed to use our package.
In the following we will highlight the main 
characteristics of QCDF90.

\section{Geometry and field definitions }

The set of all lattice sites can be subdivided into ``even'' 
and ``odd'' sites according to whether the sum of the integer 
valued coordinates
$\tt x+y+z+t$ is even or odd (checkerboard subdivision). 
There are many algorithms which
demand, especially in the context of a parallel implementation, that
even and odd variables be treated separately.
In QCDF90 we have implemented such checkerboarded separation of 
the lattice in two sublattices. Correspondingly all field variables 
are divided into even and odd variables. 
The first component of the type definition is an 
integer variable $\tt parity$ which will take
values $\tt 0$ and $\tt 1$ for variables defined over even and odd 
sites respectively.

For the gauge variables it is convenient
to include in the type a single direction 
$\mu$ component (of definite parity, of course). 
The second component of the gauge field type is an integer 
variable $\tt dir$ which takes values from $\tt 1$ 
to $\tt 4$ for variables defined in the corresponding direction.
 
In a vectorized or superscalar architecture pipelined instructions 
and longer arrays give origin to better performance. 
Therefore the lattice is most efficiently indexed by 
a single lattice volume index ranging
$\tt xyzt$ from $0$ to $NX*NY*NZ*NT/2-1$ for each sublattices
(where $Ni$ is the lattice size in direction $i$).

We have defined the following field types: 
$\tt gauge\_field$ (a $3*3$ complex matrix in a given direction);
$\tt fermi\_field$ (a $3$ component complex vector times $4$ 
spinor indices);
$\tt complex\_field$ (a complex scalar);
$\tt real\_field$ (a real scalar);
$\tt generator\_field$ ($8$ real variables in a given direction, 
for the $SU(3)$ generators).
These are defined on an even or odd sublattice. 
In addition we define the type
$\tt full\_gauge\_field$ (a collection of $8$ gauge\_field) and
the type $\tt matrix $ as a $3*3$ complex matrix.

\section{Overloaded operators}

The above type definitions are easily manipulated once one
defines overloaded operators for all possible operations 
between fields, matrices, complex and real numbers. 
The overloaded operator set includes multiplication, 
multiplication with the adjoint, division,
addition, subtraction, lambda matrices algebraic manipulations,
gamma matrices algebraic manipulations, adjoin-ing, conjugation,
real and complex traces, exponentiation, square root, contraction, 
etc...

For example, if $\tt g_i$ are gauge fields, the use of 
overloaded operators allows instructions to be as
simple as $\tt g_1=g_2*g_3$.

To implement some useful algebraic operations
on the $SU(3)$ generators we have overloaded some further operators 
which perform special operations involving 
$\tt generator\_fields$ and $\tt gauge\_fields$.  
In particular it is very important to have an efficient algorithm 
for the exponentiation of a matrix, since this operation can be a time
consuming component of several QCD calculations. The algorithm
that we have used takes advantage of the properties of the $3*3$
Hermitian traceless matrices to perform the exponentiation
with a minimal number of arithmetic operations.

\section{Shifts}

Shifts are also implemented as overloaded operators. 
For each field types C-shift implements an ordinary shift 
of the field with respect to the Cartesian geometry of the lattice.  

Moreover, because gauge theories are characterized by the
property of local gauge invariance, we find it very useful to
directly define a U-shift operator that consists of the 
shift with the appropriate parallel transport factor.
In a gauge theory the U-shift is the relevant shift operator
and is the natural building block for programming.

In QCD the efficient manipulation of the Dirac operator
is a very critical issue. Therefore for the Fermi fields we have 
also defined other shift operators which incorporate fundamental 
features of the Dirac operator.
First we define a W-shift that combines the shift with 
the parallel transport and the appropriate
gamma matrix manipulation, i.e.
acting on a Fermi field $\tt f_1$ W-shift produces a
Fermi field $\tt f_2$, given by 
\begin{equation}
f_{2,{\bf x}}= (1-\gamma_{\mu})U^{\mu}_{{\bf x}} 
f_{1,{\bf x}+\hat\mu}
\end{equation}
for positive W-shift, and
\begin{equation}
f_{2,{\bf x}}= (1+\gamma_{\mu})
U^{\dagger \mu}_{{\bf x}-\hat\mu} f_{1,{\bf x}-\hat\mu}
\end{equation}
for negative W-shift.
The direct definition of this combined operator entails 
advantages of efficiency because, from the properties of the gamma
matrices, it follows that only one half of the spin components 
undergo the transport.

We overload the (Wilson) lattice Dirac operator
and, finally, we overload the operator
Xdirac$=\gamma_5$ Dirac $\gamma_5$.

\section{Assignments}

The use of overloaded operators may imply the creation of more 
temporaries and, consequently, more motion of data than a 
straightforward implementation of operations among arrays.  
Consider for example the following operation among variables 
of type fermi\_field: $\tt f1=f1+f2+f3 $.
As far as we know, Fortran 90 does not specify how the 
variables should be passed in function calls.
As a consequence, the above instruction requires as many 
as four temporaries.
The procedure could be drastically simplified through the use of
an overloaded assignment $\tt +=$.  
The above instruction could be written $\tt f1~ += f2+f3 $
which the compiler would implement by issuing first a call to 
a function that adds $\tt f2$ and $\tt f3$ returning the result 
in $\tt t1$.  The addresses of $\tt f1$ and $\tt t1$ would then 
be passed to a subroutine that implements the operation 
$\tt f1=f1+t1$ among the components of the data types.  
The required number of copies to memory would be only two, 
instead of four.

In order to allow for these possible gains in efficiency, we 
have defined a large set of overloaded assignments.
Since Fortran 90 permits only the use of the $\tt =$ symbol 
for the assignment, we defined two global variables: a character 
variable $\tt assign\_type$ and an integer variable 
$\tt assign\_spec$ (for assign specification, introduced to 
accommodate assignments of a more elaborate nature).  
The default values of these variables are ``$\tt =$'' and 
``$\tt 0$''.  
Overloaded assignments are obtained by setting 
$\tt assign\_type$ (and possibly $\tt assign\_spec$) to the 
appropriate value immediately before the assignment. 
(Our example become $\tt assign\_type='+'; f1=f2+f3 $).

In the module ``assign\_mixed'', assignments are also defined 
between variables of different types.
For example if $\tt c_1$ is a complex\_field and 
$\tt complex$ is a complex variable, the instruction
$\tt complex=c_1$, is interpreted as setting the variable 
$\tt complex$ to the sum over all the lattice of the 
components of $\tt c_1$.

\section{Random numbers}

We have implemented a parallelizable version of the unix pseudorandom
number generator erand48, which also provides added functionality.
Erand48 is a congruential pseudorandom number generator based 
on the iterative formula
\begin{equation}
s_{i+1}=a_1*s_i+b_1 \quad {\rm mod} \; 2^{48} \ ,
\label{erand}
\end{equation}
where $a_1=\tt 0x5DEECE66D$, $b_1=\tt 0xB$, $s_i$ 
and $s_{i+1}$ are integers of at least 48 bits of precision.  
The ``seeds'' $s_i$ are converted
to real pseudorandom numbers $r_i$ with uniform distribution 
between $0$ and $1$ by $r_i=2^{-48}\, s_i$.

As presented above, the algorithm is intrinsically serial. 
However it follows from Eq.~(\ref{erand}) that the 
$\rm N^{th}$ iterate $s_{i+N}$ is still of the form 
$s_{i+N}=a_N*s_i+b_N \; {\rm mod} \; 2^{48}$ with integers
$a_N$ and $b_N$ which are uniquely determined by $a_1$, $b_1$.
The module takes advantage of this fact and of the definitions of
a global variable $\tt seeds$ to generate 
pseudorandom numbers in a parallelizable fashion.  
The module ``random\_numbers'' overloads operators that generate
Gaussian or uniformly distributed fields and all the 
necessary seed manipulations.

\section{Conditionals}

The module ``conditional''
defines six overloaded relational operators, $\tt >$, $\tt >=$,
$\tt < $, $\tt <= $, $\tt == $, $\tt /=$, and the
$\tt .Xor.$ operator.

The relational operators return a dummy logical variable and
at the same time the global variable $\tt context$ is set to 
$\tt .TRUE.$ at all sites where the relation is satisfied, 
to $\tt .FALSE.$ at all other sites. 

The operator $\tt .Xor.$ admits as operands a pair of fields
of the same type and returns a field, also of the same type,
having as elements the corresponding elements of the first operand 
at the sites where the global variable $\tt context$ is $\tt .TRUE.$, 
the elements of the second operand at the sites where 
$\tt context$ is $\tt .FALSE.$. 
This can be used to select elements out of two fields according 
to some local condition, an operation which lies at the 
foundation of stochastic simulation techniques.

\section{Precision }

To render the precision
definitions machine independent, the module ``precision''
defines two  $\tt kind $ parameters, $\tt REAL8 $ and $\tt LONG $.
These parameters store the $\tt kind $ of an 8-byte 
floating point variable and of an 8-byte integer variable.

\section{Write and read configurations}

To store and retrieve an entire SU(3) gauge field 
configuration, we developed a portable, compressed ASCII format.  
Only the first two columns of the gauge field matrices are stored, 
thanks to unitarity and unimodularity.
Our subroutines takes advantage of the fact
that all of the elements of the gauge field matrices have magnitude
smaller or equal to 1 to re-express their real and imaginary parts 
as 48bit integers.  These integers are then written in base 64,
with the digits being given by the ASCII collating sequence
starting from character ``0'' (to avoid unwanted ASCII characters).
Thus, an entire gauge field matrix is represented 
by 96 ASCII characters, without loss of numerical information.  
A detailed description of the compressing procedure is written, 
as a header, at the beginning of the stored configuration file itself.

\end{document}